\documentclass[12pt]{article}
\usepackage{epsfig}
\usepackage{graphicx}

\def\beq{\begin{equation}}
\def\eeq#1{\label{#1}\end{equation}}
\def\eeqn{\end{equation}}

\def\beqa{\begin{eqnarray}}
\def\eeqa#1{\label{#1}\end{eqnarray}}
\def\eeqan{\end{eqnarray}}

\let\bar=\overbar

\def\Dslash{\not{\hbox{\kern-4pt $D$}}}
\def\dslash{\not{\hbox{\kern-2pt $\del$}}}

\def\msb{{\bar{\ssstyle M \kern -1pt S}}}

\usepackage{fancyhdr,graphicx}
\def\Title#1{\begin{center} {\Large {\bf #1} } \end{center}}

\begin{document}

\Title{Fragmentation of Strange Quark Matter in Astrophysical Events}
% and Supernovae}

\bigskip\bigskip

%+\addcontentsline{toc}{chapter}{{\it D. Blaschke}}
%+\label{BlaschkeDavid}

\begin{raggedright}

{\it
J.E. Horvath$^{1}$~~L. Paulucci Marinho$^{2}$\\
%\thanks{\tt Email: blaschke@ift.uni.wroc.pl}
\bigskip
$^{1}$Instituto de Astronomia, Geof\'\i sica e Ci\^encias Atmosf\'ericas
Universidade de S\~ao Paulo,
R. do Mat\~ao 1226, Cidade Universit\'aria
05579-010 S\~ao Paulo, SP
Brazil\\
\bigskip
$^{2}$Universidade Federal do ABC,
Rua Santa Ad\'elia, 166,
09210-170 Santo And\'e, SP
Brazil\\
}

\end{raggedright}

\section{Introduction}

The presence of exotic quark matter among cosmic ray primaries has been a
subject of experimental and theoretical studies for many years. It is
safe to say that a milestone concerning this issue was the detection
of the very intriguing ``Centauro'' events \cite{cent}, later interpreted as
the explosion of a quark blob by Bj\"orken and Mc Lerran \cite{BML}. Over the
years, many reports have been published and some retracted \cite{ET}.
One of the latest examples of this class of events has been recently described
by Basu et al. \cite{Basu}.
The importance of these phenomena for the knowledge of hadronic physics
and its occurrence in nature can not be overstated.

It is clear that the idea of quark blobs coming from outer space needs
an appraisal in terms of production, ejection and survival of quark matter
from its original site to the detectors. Since the last 30 years a stable
form of that matter, the so-called strange quark matter (SQM) \cite{sqm} has been
studied and discussed. SQM, and its finite-$A$ version (strangelets)
is particularly suitable as a candidate for the primaries and expectations
for its direct detection (for instance, in the AMS-02 experiment \cite{experimento}
have been published elsewhere \cite{Jes}. However, there are some important
caveats concerning the very process of fragmentation of SQM (assuming it is
the true ground state of hadronic interactions), with the corresponding
consequences for the total flux. We shall discuss below some of these
issues, focusing on the physics of fragmentation. The goal would be to
check whether a high enough strangelet flux is generated, independently of
the acceleration processes \cite{MTH} within the known models widely
applied to nuclei.

\section{Strangelets from SS mergers, supernovae or both?}

Strange stars are the astrophysical realization of the SQM hypothesis. These
should mimic ordinary neutron stars in many respects, and a lot of work has
been performed to distinguish them through some clear signature. This is far
from easy, although for the moment we must assume that strange stars are
present and form shortly after a collapse/supernova event \cite{macacos}.

The possibility that the very birth of strange stars ejects some SQM into the
ISM has been considered long ago \cite{macacos}\cite{bomba}. The general picture
relies on the development of instabilities ending in a fully turbulent
propagation front \cite{yoo}, much in the same way as it is thought to happen in
ordinary thermonuclear supernovae \cite{Stan}. However, recent work \cite{Ropke}
did not find any SQM ejection, at least within the
explored physical conditions. Since (collapse) supernovae are very common,
even a tiny fraction of bulk SQM would populate the galaxy with primaries and
deserves consideration.

In a paper addressing this question, namely what happens if bulk SQM is
injected swiftly as a result of the propagation of a combustion (which
may be entangled to cause of the supernova \cite{macacos}), we \cite{VH}
have studied the fragmentation of SQM on their way out by collisions with
oxygen nuclei using a simple spallation scheme \cite{BS}. Strangelets lose
mass and kinetic energy in most cases, giving rise to a scaling law
for the final mass $m_{F}$ in terms of its initial value $m_{i}$ of the form

\begin{equation}\label{omega}
\Big(\frac{m_{F}}{m_{i}}\Big) = {\Big(\frac{\Delta_{0} \, E_{b}}{m_{p} \, E_{Oi}}\Big)}^{1/2(D-1)},
\end{equation}

where $\Delta_{0}$ is a scale for effective spallation of a fragment with that mass number,
$E_{b}$ is the binding energy released in the process, $E_{Oi}$ is the initial energy of
an oxygen nucleus and $D$ is a dimensionless quantity generally $\geq 1$ containing
the radius of the strangelet and additional quantities. The bottom line of that calculation
is that there is a real chance to obtain strangelets in the range $A \sim 10^3$ provided
the spallation with oxygen nuclei is efficient enough. Of course, this does not address
the injection mass spectrum, that is, the fragmentation prior to interactions with
the oxygen nuclei, a subject that will be considered soon.

An attempt to refine the crude model of Vucetich and Horvath \cite{VH} has been made
in the last years \cite{LauraEu}. Basically the model replaced the empirical
Boyd-Saito \cite{BS} with an abrasion-ablation scheme of the type usually
employed in the GEANT4 algorithm and similar procedures. We found that at around
$E/A \sim 10 \, MeV/A$ the spallation (abrasion) quickly dominates the fusion and
scattering cross-sections, giving rise to a much moderate downsizing of
simulated strangelet-nitrogen collisions. Therefore, it is likely that the first
models overestimated the reprocessing of strangelets and $A \sim 10^{2} - 10^{3}$
fragments could be produced only if the initial baryon number was not very high. Thus,
the initial mass distribution becomes an essential ingredient.

The idea of fragmentation of bulk ejected SQM itself was never considered in depth
because the binding energy of the former always increases with baryon number $A$.
At first sight, producing fragments out of a big chunk would go against energetics
because the process must work ``uphill''. Madsen \cite{Jes2} equated the gravitational
energy to the surface energy to estimate that around $A \sim 10^{38}$ the SQM should
break up in chunks of that $A$. His picture of big nuggets orbiting a SS lead him
to identify orbital collisions as the mechanism to further fragment the chunks
into $A \sim 10^{3}$ strangelets, that is, a very substantial reprocessing indeed.

In complete analogy with bulk nuclear matter, it is expected, however, that a consideration
of the free energy of the system (not just the binding energy) indicates the fragmentation
point and also the mass distribution of the fragments. The physical situation of any
ejected SQM in bulk corresponds to an expansion and cooling of the system, and therefore
excluded volume at the fragmentation point and vacuum ``melting'' should be
important as well. Note that this is independent of reprocessing by collisions. Moreover,
such a situation would be encountered in the first moments after a supernova shock
breakout, but also in the situation of ejection at the merger of two compact strange stars
\cite{merging}. The latter process could eject as much as $10^{-4} M_{\odot}$ in the
form of SQM, but this would produce either planetary mass chunk(s) or require very
efficient reprocessing, as explained above, to later match the reported CR primary range.
We shall now present the results of a calculation intended to explain the fragmentation
and the mass distribution of the fragments, without considering further collisions or
acceleration.

\section{Multifragmentation of SQM}

Statistical multifragmentation models are known to
reproduce the fragment distribution of ordinary
heavy nuclei and can be applied whenever the
temperature $T$ is of the order of the binding energy.
In the grand canonical ensamble, the partition function reads \cite{Bugaev}

\begin{equation}\label{omega}
\omega_A=V\Big(\frac{mTA}{2 \pi}\Big)^{3/2}e^{-f_A/T},
\end{equation}

with $f_A$ the internal free energy of the fragment

\begin{equation}\label{free_energy}
f_A=-WA+\sigma A^{2/3}+CA^{1/3},
\end{equation}

representing the bulk, surface and curvature terms respectively.
Performing the Fourier transform of eq.(2) $\mathcal{Z}$, and recalling that

\begin{equation}
p(T,\mu)=T\lim_{V \to \infty}\frac{\ln \mathcal{Z}(V,T,\mu)}{V},
\end{equation}

the problem of the fragmentation is reduced to finding the singularities of
the isobaric partition function, yielding the gas and liquid pressures as

\begin{eqnarray}
p_g(T,\mu)&=&T\Big(\frac{mT}{2\pi}\Big)^{3/2}\Big\{z_1e^{\frac{\mu-bp_g}{T}}+
\sum_{A=2}^{\infty}A^{3/2}e^{[(\nu-bp_g)A-\sigma A^{2/3}-CA^{1/3}]/T}\Big\}, \\
p_l(T,\mu)&=&\frac{\nu}{b},
\end{eqnarray}
\\
where $\nu=\mu+W$ is the (shifted) chemical potential. The fragmentation
spectrum is obtained from the derivatives of the pressure assuming
chemical equilibrium. However, when these quantities were calculated
for SQM within the MIT bag model framework and in the
color-flavor-locked (CFL) state \cite{Alf}, the surprising result that the peak of the
fragments happens in the region $A \ll 1$ was obtained in all cases, even
introducing a simple Van der Waals approximation. This may be due to
the hidden assumption about a constant energy of the vacuum (represented by
an MIT value $B$. However, physical considerations can be made in favor
of a ``melting'' of the vacuum picture: a fraction of the vacuum energy is
being used to create the surfaces and curvature energies, therefore there should
be an energy density difference between the phases

\begin{eqnarray*}
W_l=W_0+Bv_{liq}, \\
W_g=W_0+Bv_{gas}.
\end{eqnarray*}

\begin{figure}[]
\includegraphics[width=0.7\textwidth]{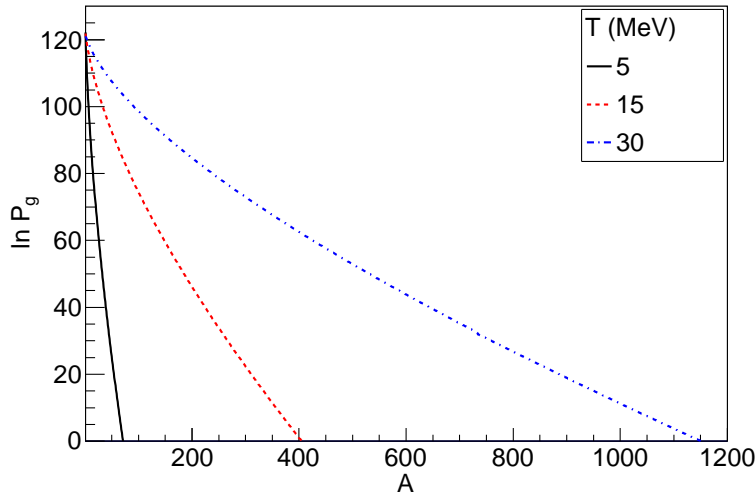}
\caption{The distribution function of the fragments (strangelets) normalized to $10^{-4} M_{\odot}$ for 
three different values of the temperature (indicated). Fragmentation at higher temperatures 
favor a tail of heavier particles, but also destabilizes the small $A$ strangelets. As a result, 
the net number of surviving strangelets decreases at higher temperatures.}
\label{fig:Norm}
\end{figure}

After this correction, the distribution shifts to higher values of $A$.
We have normalized the probabilities to a total of $10^{4} M_{\odot}$
as expected from the above considerations \cite{PH}. The masses of the fragments
(strangelets) reach higher values when progressively higher
temperatures are considered (Fig. ~\ref{fig:Norm}). However, as is well-known
the stability of the strangelets diminishes with rising temperature, and
an increasing number of fragments should decay into ordinary hadrons

\section{Conclusions}

From the calculations performed in a series of studies we may
conclude that either SQM does not like to fragment, or if it does
an overwhelmingly large fraction of the original baryon number
would be ultimately go into ordinary particles, not strangelets.
A similar calculation by Biswas et al \cite{biswas} has pointed
out the sensitivity to the treatment of the binding energy dependence,
while the results above become odd without a substantial
role of the vacuum melting. If the fragmentation destroys most
of strangelets, the flux of exotics in CRs could be minuscule.
Actually, the (optimistic) expression for the latter given by Madsen
\cite{Jes3}

\begin{equation}\label{flux}
F = 5 \, \times \, 10^{5} (m^{2} \, yr \, sterad)^{-1} \, \times \, R_{-4} \, \times \, M_{-2}
\, \times \, V_{100}^{-1} \, \times \, t_{7} \, ,
\end{equation}

with $R_{-4}$ the rate of mergers in units of $10^{-4} yr^{-1}$, $M_{-2}$ the ejected mass going
into strangelets in units of $10^{-2} M_{\odot}$ should be multiplied by an efficiency factor
of at least $10^{-5}$ representing the strangelet fragility under decays. This is independent of
the possibility that the ejection could indeed be $zero$ (\cite{merging}) in these events.
Adding the caveats already made in the supernova case, we conclude that if strangelets are
positively detected after all, the fragmentation should proceed out of equilibrium, a
possibility yet to be explored.

\subsection*{Acknowledgements}

JEH expresses his thanks to the organizers of the CSQCD IV conference for the
realization of a live and productive event in which many ideas and paths
were discussed and suggested to him. The financial support of the Fapesp
(S\~ao Paulo State) and CNPq (Brazil) agencies is gratefully
acknowledge by both authors.

\end{document}